\documentclass[aps,preprint,prd,nofootinbib]{revtex4}
\usepackage{graphicx}
\usepackage{psfrag}

\begin{document}

\title{The Productions and Strong Decays of $D_q(2S)$ and $B_q(2S)$}
\author{Zhi-Hui Wang$^{[1]}$, Guo-Li Wang$^{[1,2]}$\footnote{gl\_wang@hit.edu.cn}, Jin-Mei Zhang$^{[3]}$, Tian-Hong Wang$^{[1]}$\\}
\address{$^1$Department of Physics, Harbin Institute of
Technology, Harbin, 150001, China\\$^2$PITT PACC, Department of
Physics $\&$ Astronomy, University of Pittsburgh, PA 15260, USA
\\$^3$Xiamen Institute of
Standardization, Xiamen 361004, China}

 \baselineskip=20pt

\begin{abstract}
 We study the productions of first radial excited states
$D_q(2S)$ ($q=u,~d,~s$) and $B_q(2S)$ in exclusive semi-leptonic
$B_{q'}$ ($q'=u,~d,~s,~c$) decays by the improved Bethe-Salpeter
method. These $2S$ states can be detected through their strong
decays to ground mesons, where the strong decays are calculated by
the low energy approximation as well as the impulse approximation.
Some channels have ratios of order $10^{-4}$: $Br(B^+\to\bar
D^0(2S)\ell^+{\nu_\ell})\times Br(\bar D^0(2S)\to \bar
D^{*}\pi)\approx(4.9\pm4.0)\times10^{-4}$, $Br(B^0\to
D^-(2S)\ell^+{\nu_\ell})\times Br(D^-(2S)\to \bar
D^{*}\pi)\approx(4.4\pm3.4)\times10^{-4}$. These channels could be
measured by the current B-factories. For $D_s(2S)$, we also obtain
a relative large ratio: $Br(B_s^0\to
D_s^-(2S)\ell^+{\nu_\ell})\times Br(D_s^-(2S)\to \bar D^{*}\bar
K)\approx (9.9\pm14.9)\times10^{-4}$. Although there are
discrepancies of the full decay width between the theoretical
predictions of ${D}^0(2S)$ and experimental results of
$D(2550)^0$, the new detected state $D(2550)^0$ is very likely the
${D}^0(2S)$ state. \vspace*{0.5cm}


\end{abstract}

\maketitle
\section{Introduction}

In recent years, great progress has been made in hadronic mass
spectra. There are many new states that have been observed, e.g.,
the new particles $D^{*}_{s0}(2317)$ \cite{2317}, $D_{s1}(2700)$,
$D_{sJ}(2860)$ \cite{2700}, $\eta'_c$ \cite{etac}, $X(3872)$
\cite{3872}, $X(3940)$ \cite{x3940}, $Y(3940)$ \cite{y3940},
$Z(3930)$ \cite{z3930} and $Y(4260)$ \cite{y4260}. Some of these
new states are $P$-wave ($L=1$) state candidates, such as
$D^{*}_{s0}(2317)$, and some of them are first radial excited 2$S$
($L=0$, $n=2$) state candidates, e.g., $D_{s1}(2700)$ and
$\eta'_c$. Recently, other new 2$S$ state candidates $D(2550)^0$
and $D^*(2600)^0$ are observed in inclusive $e^+e^-$ collisions
near $\sqrt s=10.58$ GeV \cite{2550}.

Besides the progress in experiment, there are many approaches to
study the heavy excited states in theory, e.g., the authors of
\cite{Ebert} considered $B_c$ decays to excited $2P$ and $3S$
charmonium states with the relativistic quark model; the authors
of \cite{Lu} calculated the decay of $B_c\to X_{c\bar c}l\nu_l$
where $X_{c\bar c}$ was an excited charmonium state with the
light-cone QCD sum rules approach; the authors of \cite{4612} used
generalized factorization together with $SU(3)_F$ symmetry to
predict the branching ratios of $B_s\to M_{c\bar c}+L$ where
$M_{c\bar c}$ was a charmonium state and $L$ was a light scalar;
using the ISGW2 quark model, and the authors of \cite{non} studied
$B_c\to X_{c\bar c}M$ decays, where $X_{c\bar c}$ was a radial
excited charmonium $\eta_c(2S)$ or $\psi(2S)$.

Although several papers considered the topics of radial excited
$2S$ states \cite{isgur,scora,Ebert2,leibovich}, there is still
lack of theoretical investigation for the radial excited states
$D_q(2S)$ or $B_q(2S)$, where $q$ is a light quark. One may also
note that there is no other heavy-light 2$S$ state, which has ever
been confirmed by the experiment except charmonium and
bottomonium, this means we have little knowledge about heavy-light
$2S$ states, so the study of heavy-light $2S$ states will enlarge
our knowledge of bound states and QCD.

There are many methods to detect the heavy-light 2$S$ states
experimentally. For example, by analyzing the inclusive
productions of $D^+\pi^-$, $D^0\pi^+$ and $D^{*+}\pi^-$ systems,
the Babar collaboration found new mesons like $D(2550)^0$ and
$D^*(2600)^0$ \cite{2550}. Since there are a large number of $B$
data in two B-factories, and the LHC will produce large data of
$B_s$ and $B_c$, there will be a best and convenient way to detect
heavy-light 2$S$ states by $B_{q'}$ exclusive decays. In theory,
the properties of the mesons were studied by many approaches
together with the Dyson-Schwinger(DS) equation of QCD or the
Bethe-Salpeter(BS) equation or both of them
\cite{Tnguyen,Akrassnigg,Souchlas}. In this paper, we will study
the productions of heavy-light 2$S$ states in exclusive
semi-leptonic decays of $B_{q'}$ mesons by the instantaneous
approximate BS method \cite{mass, chen}.

Knowing mass and life time (or full width) is helpful to detect
the resonance experimentally. In a previous letter
\cite{zhangwang}, we have studied mass and strong decay of some
$D_q(2S)$ and $B_q(2S)$ states by the BS method. Unfortunately,
there are large mass and width discrepancies between our predicted
result of
 $D^0(2S)$ and that of $D(2550)^0$, which is the candidate of
$D^0(2S)$ in the experiment \cite{2550}. Our predicted mass is
$2.39$ GeV of $D^0(2S)$ \cite{zhangwang}, while Babar's result is
$2539.4\pm4.5\pm6.8$ MeV of $D(2550)^0$ \cite{2550}. The total
strong decay width of our calculation
 is $8.4$
MeV (note that we missed a parameter $0.5$ in all the strong
decays whose final state involve $\pi^0$), which is much smaller
than the experimental value $130\pm12\pm13$ MeV. There are several
theoretical approaches to study the  strong decays of this new
state \cite{zhangwang,wangzg,liu,zhongxh,zhangal,lidm}. We show
the theoretical estimated full decay widths and the experimental
data in TABLE~\ref{D2550}. One can see that there are large
discrepancies between experimental and theoretical results except
the result of \cite{zhangal}.

\begin{table*}
\centering  \caption{\label{D2550}\small The total decay widths
(in unit of MeV) of $D(2550)^0$ treated as the $D(2S)^0$ state;
our results are estimated by the low energy approximation and the
impulse approximation (in parentheses).} \vspace{5mm}
\doublerulesep2pt
\begin{tabular}{ccccccc}
\hline

\hline
 Ex~\cite{2550}&this
Paper&\cite{zhangwang}&~\cite{liu}&~\cite{zhongxh}&~\cite{zhangal}&~\cite{lidm}
\\ \hline
$130\pm12\pm13$&$43\pm23(47\pm17)$&8.4&8&22.1&124.1&45.35
\\
\hline

\hline
\end{tabular}
\end{table*}

From our previous calculations, we find that there are three main
reasons that result in discrepancies. First, we have chosen a
simple potential--the Cornell potential--in order to reduce the
difficulty of solving the BS equation, which is very complicated
in this work. Second, we chose a group of old input parameters.
Since there was no information of the 2$S$ state mass in the
previous letter \cite{zhangwang}, we obtained the masses of 2$S$
states by fitting data of ground states with old parameters:
$m_b=5.224$ GeV, $m_c=1.7553$ GeV, $m_s=0.487$ GeV, $m_u\simeq
m_d=0.3$ GeV, and other parameters that character the potential
\cite{w1}. Recently, by fitting data of charmonia and bottomonia
which include higher excited states, and combining with the
results of decay constants, we give new set of parameters:
$m_b=4.96$ GeV, $m_c=1.62$ GeV, $m_s=0.5$ GeV, $m_u\simeq m_d=0.3$
GeV \cite{mass}. Third, our results are model dependent and we
only consider two OZI-allowed channels to estimate the full decay
width. Furthermore, the theoretical prediction of decay width is
very sensitive to the mass of $2S$ state.

In this paper, we focus on the decay and production of the $2S$
states, not on the mass spectra, so we can vary the free parameter
$V_0$ to obtain the new mass spectra in TABLE \ref{masses} and the
numerical values of wavefunctions, which are used to calculate the
transition matrix elements in appendix B. By varying all the input
parameters simultaneously within 5\% of central values, we obtain
the uncertainties of masses in TABLE \ref{masses}.

Although we focus on the production of heavy-light $2S$ states in
the semi-leptonic decays of $B_{q'}$, we would like to
re-calculate their strong decays by the re-predicted mass spectra.
 The strong decay widths are very sensitive to the
kinematic range, so some new strong decay channels with higher
mass of $2S$ states can exist, e.g., there is a strong decay with
a $P$-wave state involved in final states. Finally, we calculate
the ratios of strong decays to reduce the effect of model
dependence, and estimate the production ratios of $B_{q'}$
semi-leptonic decay to the first radial excited states, which are
reconstructed by the ground particles in terms of strong decay.

The rest of the paper is organized as follows. In section II, we
show the formulations of semi-leptonic and strong decays. We give
the hadronic matrix elements of semi-leptonic and strong decays in
section III. The results and discussions are given in section IV.
In the appendices we introduce BS equation and give some necessary
formulas for the calculations of semi-leptonic and strong decays.

\section{The formulations of semi-leptonic decays and strong
decays} In this section, we present the formulations of $B_{q'}$
mesons semi-leptonic decay to $2S$ mesons and the strong decays of
 $2S$ mesons.
\subsection{Semi-leptonic decay of $B_{q'}$}

\begin{figure}
 \centering
\includegraphics[width=3.5in]{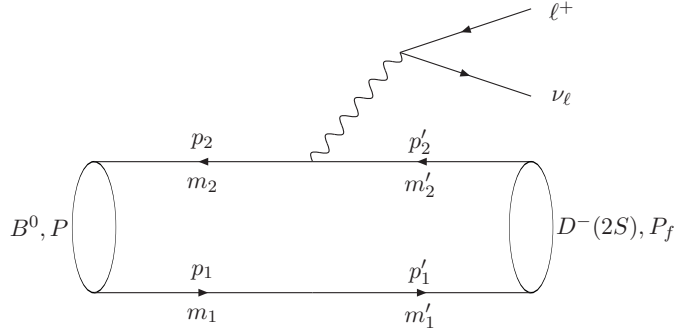}
\caption{\label{semi} Feynman diagram of the semi-leptonic
decay $B^0\to D^-(2S)\ell^+\nu_\ell$.}
\end{figure}

As we mentioned previously, the best way to study 2$S$ state is
through the exclusive semi-leptonic decay of initial heavy meson
($B^0$, $B^+$, $B_s^0$ or $B_c$). Here, we take the $B^0\to
D^-(2S)\ell^+{\nu_\ell}$ (see figure \ref{semi}) as an example to
show the formulation. The amplitude of this process is
\begin{eqnarray}\label{T}
T=\frac{G_F}{\sqrt{2}}V_{cb}\bar{u}_{\nu_\ell}\gamma^{\mu}(1-\gamma_5)
v_{\ell}\langle D^-_{2S}(P_f)|J_{\mu}|B^0(P)\rangle\,,
\end{eqnarray}
where $V_{cb}$ is the CKM matrix element,
$J_{\mu}=V_{\mu}-A_{\mu}$ is the charged weak current, and $P$ and
$P_f$ are the momentum of the initial meson $B^0$ and the final
state $D^-(2S)$, respectively. The hadronic part can be written as
\begin{eqnarray}\label{form}
&&\langle
D^-_{2S}(P_f)|V_{\mu}|B^0(P)\rangle=f_+(P+P_f)_{\mu}+f_-(P-P_f)_{\mu},
\nonumber\\
 &&\langle D^-_{2S}(P_f)|A_{\mu}|B^0(P)\rangle=0,
\end{eqnarray}
where $f_+, f_-$ are the Lorentz invariant form factors.

We define $x\equiv E_\ell/M,\;\; y\equiv (P-P_f)^2/M^2$, where
$E_\ell$ is the energy of the final charge lepton and $M$ is the
mass of initial meson. The differential width of the decay can be
reduced to
\begin{eqnarray}\label{differ}
&&\frac{d^2\Gamma}{dxdy}=|V_{bc}|^2\frac{G_F^2M^5}{64{\pi}^3}
\nonumber\\
&&\left\{{\beta}_{++}\left[4\left(2x(1-\frac{M_f^2}{M^2}+y)-4x^2-y\right)
+\frac{m_\ell^2}{M^2}\left(8x+4\frac{M_f^2}{M^2}-3y-\frac{m_\ell^2}{M^2}\right)\right]\right.\nonumber\\
&&\left.({\beta}_{+-}+{\beta}_{-+})\frac{m_\ell^2}{M^2}
\left(2-4x+y-2\frac{M_f^2}{M^2}+\frac{m_\ell^2}{M^2}\right)
+{\beta}_{--}\frac{m_\ell^2}{M^2}\left(y-\frac{m_\ell^2}{M^2}\right)\right\}\,,
\end{eqnarray}
where $M_f$, $m_\ell$ are the masses of the meson and the lepton in
final, respectively. $\beta_{++}=f^2_+$,
$\beta_{+-}=\beta_{-+}=f_+f_-$, $\beta_{--}=f^2_-$.

\subsection{Strong decay of $2S$ mesons }

\begin{figure}
 \centering
\includegraphics[width=3.5in]{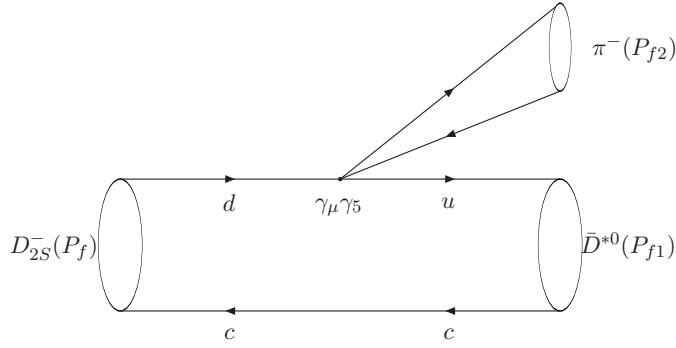}
\caption{\label{fig.strong} Feynman diagram of strong decay
$D^-(2S)\to \bar D^{*0}\pi^-$ (low-energy approximation).}
\end{figure}

As an example, we consider the OZI-allowed strong decay
$D^-(2S)\to \bar D^{*0}\pi^-$ (see figure \ref{fig.strong}). In
this work, we take the instantaneous approximation for the
interaction kernel in meson; it is fit to describe the double
heavy mesons ($B_c, \eta_c$) and heavy-light mesons ($D_q$, $B_q$)
\cite{BsDs,chris}, but it is inapplicable to double light meson
($K$, $\pi$), which have complicated internal structure. In this
work, we take the reduction formula, PCAC relation and low energy
theorem to deal with the strong decay as we did in \cite{Chang}.
The strong decay amplitude of figure \ref{fig.strong} can be
written as \cite{zhangwang,Chang}
\begin{equation}\label{strong}
T=\frac{P_{f_2}^\mu}{f_{P_{f_2}}}\langle \bar
D^{*0}(P_{f1})|\bar{u}\gamma_\mu\gamma_5 d|
D^-_{2S}(P_f)\rangle,
\end{equation}
where $P_f$, $P_{f_1}$ and $P_{f_2}$ are the momenta of the
 $D^-(2S)$, final states $\bar D^{*0}$ and $\pi^-$,
respectively, $f_{P_{f_2}}$ is the decay constant of $\pi^-$; we
call this method the low energy approximation in this paper.

\begin{figure}
 \centering
\includegraphics[width=3.5in]{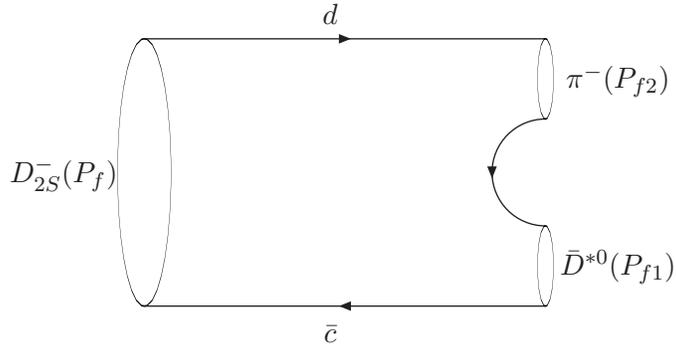}
\caption{\label{fig-impulse} Feynman diagram of strong decay
$D^-(2S)\to \bar D^{*0}\pi^-$ (impulse approximation).}
\end{figure}

As a comparison, we also calculate the strong decays using an
alternative method called the impulse approximation
\cite{cdroberts1,cdroberts2}. According to this method, the decay
amplitude of $D^-(2S)\to \bar D^{*0}\pi^-$ can be written as (see
Fig.~\ref{fig-impulse})
\begin{eqnarray}\label{amplitude-impulse}
&&T=\int\frac{d^4q_{f}d^4q_{f1}}{(2\pi)^4}\nonumber\\
&&
Tr\left\{S_1(p'_1)\eta_{P_f}(q_f)S_2(p'_2)\bar\eta_{P_{f1}}(q_{f1})S'_1(p''_1)
\Gamma_\pi(P_{f2})\delta^4(\alpha'_2P_f-q_f-(\alpha''_2P_{f1}-q_{f1}))\right\},
\end{eqnarray}
where $q_f$, $q_{f1}$ are the relative momentum of quark-anti-quark in $D^-(2S)$ and $\bar D^{*0}$,
 $S_1(p'_1)$, $S_2(p'_2)$ and $S'_1(p''_1)$ are propagators, $\eta_{P_f}(q_f)$,
$\bar\eta_{P_{f1}}(q_{f1})$ are the heavy-meson BS amplitudes,
respectively; $p'_1=\alpha'_1P_f+q_f$; $p'_2=\alpha'_2P_f-q_f$;
$p''_1=\alpha''_1P_{f1}+q_{f1}$;
 $\alpha'_1=\frac{m_d}{m_d+m_c}$; $\alpha'_2=\frac{m_c}{m_d+m_c}$;
 $\alpha''_1=\frac{m_u}{m_u+m_c}$; and $\alpha''_2=\frac{m_c}{m_u+m_c}$.
 $\Gamma_\pi(P_{f2})$ is the BS amplitude of $\pi$.

 After instantaneous approximation, equation (\ref{amplitude-impulse}) can be written as
 \begin{eqnarray}\label{impulse-2}
T=\int\frac{d^3q_{f}}{(2\pi)^3}
Tr\left\{\varphi^{++}_{P_f}(q_{f\bot})\frac{\not\!P_f}{M_f}\bar\varphi^{++}_{P_{f1}}(q_{f1\bot})
\Gamma_\pi(q_{f\bot}; P_{f2})\right\},
\end{eqnarray}
where $q_{f\bot}=(0, \vec q_f)$,
${q}_{{f1\bot}}={q}_{f\bot}+\frac{m_c}{m_c+m_u}{P}_{f1\bot}$,
${\varphi}^{++}_{{P_f}}({{q}_{f\bot}})$ is the positive energy
wavefunction for $D^-_{2S}(P_f)$.
${\varphi}^{++}_{_{P_{f1}}}({q}_{_{f1\bot}})$ is the positive
energy wavefunction for $\bar D^{*0}$. $\Gamma_\pi(q_{f\bot};
P_{f2})=i\gamma_5\frac{\sqrt{2}}{f_{\pi}}B_\pi(q_{f\bot}^2)$; one
can find detailed calculation of $B_\pi(q_{f\bot}^2)$ in
\cite{cdroberts1,cdroberts2,constant}.

There are two channels for the $D^-(2S)$ meson: $D^-(2S)\to \bar
D^{*0}\pi^-$ ($0^-(2S)\to 1^-0^-$) and $D^-(2S)\to
 \bar
D^{*}_{0}(2400)^0\pi^-$ ($0^-(2S)\to 0^+0^-$).
 The strong decay amplitudes can be described by the strong
coupling constants
\begin{eqnarray}\label{constant}
&&T(D^-(2S)\to
\bar D^{*0}\pi^-)=G_{D^-(2S)\bar D^{*0}\pi}(\varepsilon^{(\lambda)}\cdot P_f),\nonumber\\
&&T(D^-(2S)\to \bar D^{*}_{0}(2400)^0\pi^-)=G_{D^-(2S)\bar
D^{*0}_{0}\pi},
\end{eqnarray}
where $G_{D^-(2S)\bar D^{*0}\pi}$ and $G_{D^-(2S)\bar
D^{*0}_{0}\pi}$ are the strong coupling constants, and
 $\varepsilon$ is the polarization vector of meson $\bar D^{*0}$.

With equation (\ref{constant}), we obtain the decay widths
\begin{eqnarray}\label{strongwidth}
&&\Gamma_{D^-(2S)\bar D^{*0}\pi}=\frac{|\vec{P}_{f1}|}{8\pi M_f^2}\sum_{\lambda}|T(D^-(2S)\to
\bar D^{*0}\pi^-)|^2,\nonumber\\
&&\Gamma_{D^-(2S)\bar
D^{*}_{0}(2400)^0\pi^-}=\frac{|\vec{P}_{f1}|}{8\pi
M_f^2}|T(D^-(2S)\to \bar D^{*}_{0}(2400)^0\pi^-)|^2.
\end{eqnarray}

From equations (\ref{T}) and (\ref{strong}),
 we find that the main task of semi-leptonic and strong decay is
 to calculate the amplitudes
 $\langle D^-_{2S}(P_f)|J_{\mu}|B^0(P)\rangle$ and $\langle \bar
D^{*0}(P_{f1})|\bar{u}\gamma_\mu\gamma_5 d|
D^-_{2S}(P_f)\rangle$.
\section{The hadronic matrix elements of semi-leptonic decay and strong decay}

\subsection{Hadronic matrix element of semi-leptonic decay}
The calculation of the hadronic matrix element are different from
model to model. In this paper, we combine the BS method that is
based on the relativistic BS equation with the Mandelstam
formalism \cite{Mand} and relativistic wavefunctions to calculate
the hadronic matrix element. The numerical values of wavefunctions
have been obtained by solving the full Salpeter equation that we
have introduced in the appendices. As an example, we consider the
semi-leptonic decay $B^0\to D^-(2S)\ell^+{\nu_\ell}$. In this way,
at the leading order, the hadronic matrix element can be written
as \cite{chen}
\begin{eqnarray}\label{matrix}
\langle D^-_{2S}(P_f)|J_{\mu}|B^0(P)\rangle=
\int\frac{d{\vec{q}}}{(2\pi)^3}{\rm Tr}\left[
\bar{\varphi}^{++}_{_{P_{f}}}(\vec {q}_{_f})\frac{\not\!P}{M}
{\varphi}^{++}_{_P}({\vec{q}})\gamma_{\mu}(1-\gamma_5)\right]\, ,
\end{eqnarray}
where $\vec{q}$ ($\vec{q}_{_f}$) is the relative three-momentum
between the quark and the anti-quark in the initial (final) meson
and $\vec{q}_{_f}=\vec{q}-\alpha'_1{\vec{r}}$. $M$ is the mass of
$B^0$, ${\vec{r}}$ is the three-dimensional momentum of $D^-(2S)$,
${\varphi}^{++}_P(\vec q)$ is the positive Salpeter wavefunction
of $B^0$ meson and ${\varphi}^{++}_{P_f}(\vec q_f)$ is the
positive Salpeter wavefunction of $D^-(2S)$ meson,
$\bar{\varphi}^{++}_{_{P_f}}=\gamma_0({\varphi}^{++}_{_{P_f}})^{\dagger}\gamma_0$.
We have given the Salpeter wavefunctions for the different mesons
and form factors in appendix B.

\subsection{Hadronic matrix element of strong decays}

We have obtained the amplitude of strong decay $D^-(2S)\to \bar
D^{*0}\pi^-$ in equation (\ref{strong}), and the key factor is to
calculate the hadronic matrix element $\langle \bar
D^{*0}(P_{f1})|\bar{u}\gamma_\mu\gamma_5 d| D^-_{2S}(P_f)\rangle$.
The hadronic matrix element $\langle \bar
D^{*0}(P_{f1})|\bar{u}\gamma_\mu\gamma_5 d| D^-_{2S}(P_f)\rangle$
can be written as \cite{Chang}
\begin{eqnarray}\label{a08}
\langle \bar D^{*0}(P_{f1})|\bar{u}\gamma_\mu\gamma_5 u|
D^-_{2S}(P_f)\rangle =
\int\frac{d{\vec{q}_{_f}}}{(2\pi)^3}Tr\left[
\bar{\varphi}^{++}_{_{P_{f1}}}(\vec{q}_{_{f1}})\gamma_{\mu}\gamma_5
{\varphi}^{++}_{_{P_f}}({\vec{q}_{_{f}}})\frac{\not\!P_f}{M_f}
\right].
\end{eqnarray}
 We give the relation of wavefunctions and the strong coupling constants of different strong decays in appendix B.

\section{Number results and discussions}

\subsection{Semi-leptonic decays}
\begin{figure}
\centering
\includegraphics[height=5cm]{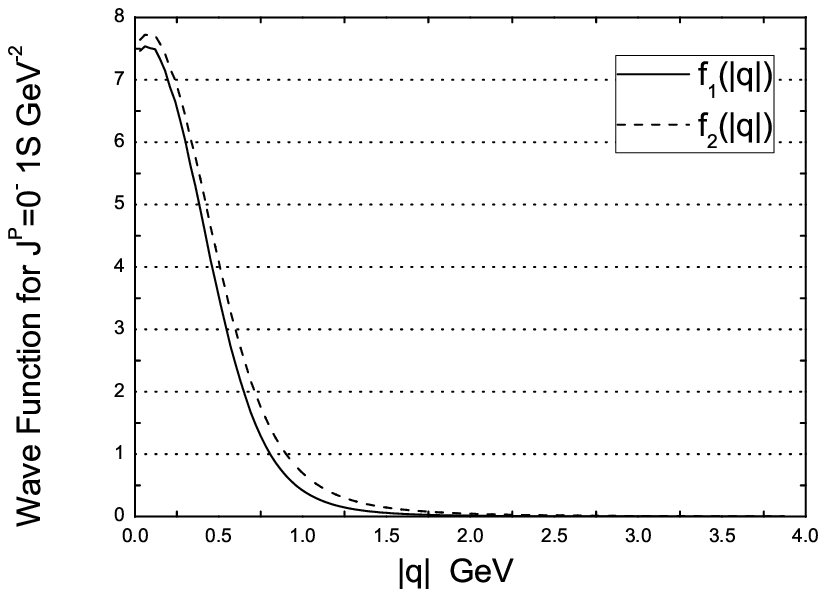}
\includegraphics[height=5cm]{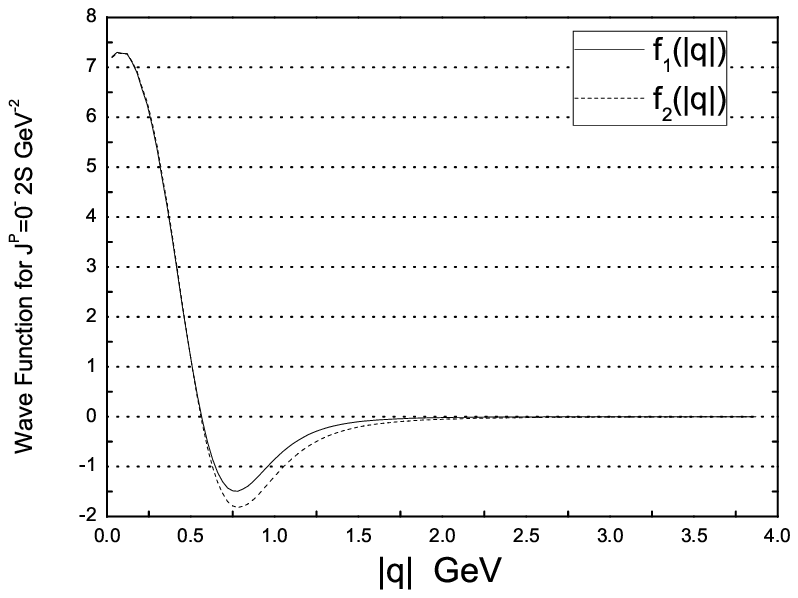}
\caption{\label{wavefunction}The wavefunctions of $1S$ and the
$2S$ of $D^-$ meson with $J^P=0^-$.}
\end{figure}

In order to fix the Cornell potential in equation (\ref{eq16}) and
masses of quarks,
 we take these parameters: $a=e=2.7183,
\lambda=0.21$ GeV$^2$, ${\Lambda}_{QCD}=0.27$ GeV, $\alpha=0.06$
GeV, $m_b=4.96$ GeV, $m_c=1.62$ GeV, $m_s=0.5$ GeV, $m_u=0.305$
GeV, $m_d=0.311$ GeV, $etc$ \cite{mass}, which are best to fit the
mass spectra of ground states $D_q$, $B_q$ and other heavy mesons.
In the previous letter \cite{zhangwang}, we have obtained masses
of ground states and 2$S$ states by solving the BS equation. But
we find that the mass of $D^0(2S)$ in \cite{zhangwang} is smaller
than the one of $D(2550)^0$. In order to fit the experimental
value of $D(2550)^0$, we change the free parameter $V_0(D^0)$ to
$V_0(D^0(2S))$. For other heavy-light 2$S$ states that do not have
experimental data, we vary
 $V_0(1S)$ that has the same value as
 $\Delta V_0=V_0(D^0(2S))-V_0(D^0)$ to obtain the masses spectra of other heavy-light 2$S$
states. Then, we obtain the mass spectra of different mesons in
TABLE \ref{masses}. Varying all the input parameters($\lambda$,
$\Lambda_{QCD}$, $etc$) simultaneously within 5\% of the central
values, we obtain the uncertainties of masses. To show the
numerical results of wavefunctions explicitly, we plot the
wavefunctions of $1S$ and $2S$ states for $D^-$ meson with
$J^P=0^-$ in figure \ref{wavefunction}. For semi-leptonic decays,
we need to input the CKM matrix elements:
 $V_{cb}=0.0406$, $V_{cd}=0.23$, $V_{cs}=0.97334$ and the lifetimes of mesons: $\tau_{B^0}=1.53\times10^{-12}s$,
$\tau_{B^+}=1.638\times10^{-12}s$, $\tau_{B_c} =
0.453\times10^{-12}s$, $\tau_{B_s^0} = 1.47\times10^{-12}s$,
which are taken from PDG \cite{PDG}.

\begin{table*}
\centering  \caption{\label{masses}\small Masses of the 1$S$ and
2$S$ states (in unit of MeV), `Ex' means the experimental data
from PDG \cite{PDG} and \cite{2550}, `Th' means our prediction.}
\vspace{5mm} \doublerulesep2pt
\begin{tabular}{cccccc}
\hline

\hline &Th&Ex& &Th&Ex
\\ \hline
$D^-(1S)$&1869.4&1869.6$\pm0.16$&$~\bar{D}^0(1S)$&1865.0&1864.83$\pm0.14$
\\
$D^-(2S)$&$2560\pm110$&
&$~\bar{D}^0(2S)$&$2550\pm109$&$2539.4\pm4.5\pm6.8$ \cite{2550}
\\
$M(2S)-M(1S)$&$691\pm110$&
&$M(2S)-M(1S)$&$685\pm109$&$674.6\pm4.5\pm6.8$
\\ \hline
$B^0(1S)$&5279.5&5279.5$\pm0.3$ &$D_s^-(1S)$&1968.2&1968.47$\pm0.33$
\\
$B^0(2S)$&$5930\pm279$&     &$D_s^-(2S)$&$2641\pm123$&
\\
$M(2S)-M(1S)$&$651\pm279$&    &$M(2S)-M(1S)$&$673\pm123$&
\\ \hline
$B^{+}(1S)$&5279.0&5279.17$\pm0.29$&$B_s^0(1S)$&5367.9&5366.3$\pm0.6$
\\
$B^{+}(2S)$&$5930\pm280$&&$B_s^0(2S)$&$6020\pm281$&
\\
$M(2S)-M(1S)$&$651\pm280$&&$M(2S)-M(1S)$&$652\pm281$&
\\ \hline

\hline
\end{tabular}
\end{table*}
\begin{figure}
 \centering
 \includegraphics[height=5cm]{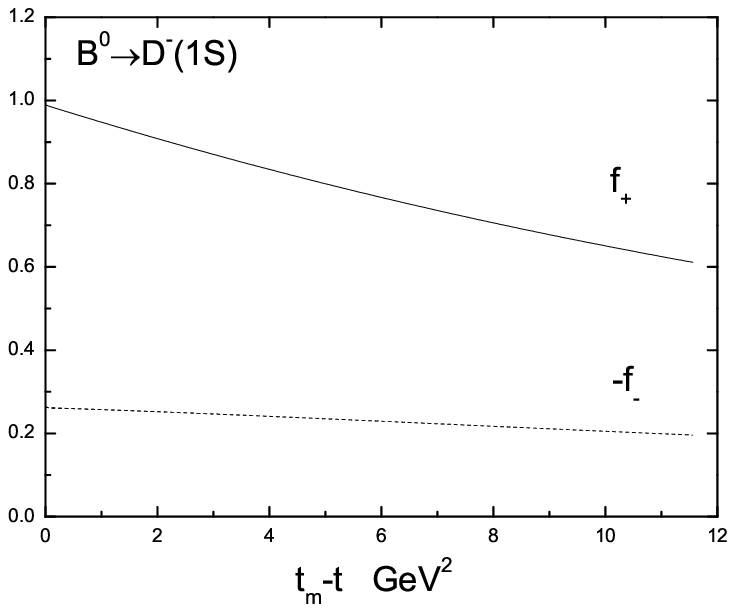}
\includegraphics[height=5cm]{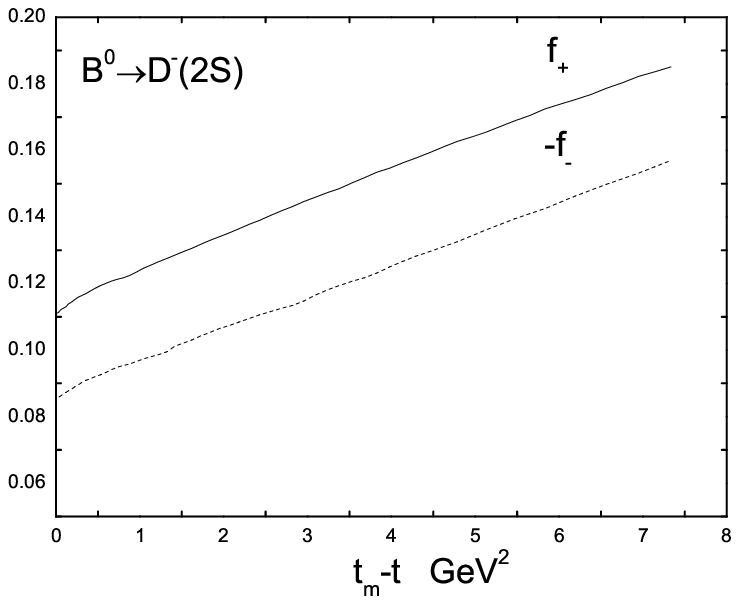}
\caption{\label{figform}The form factors of $B^0\to D^-(1S)\ell^+\nu_\ell$
 and $B^0\to D^-(2S)\ell^+\nu_\ell$. }
\end{figure}
\begin{table*}
\centering  \caption{\label{widthbr}\small The decay widths and
branching ratios of exclusive semi-leptonic decay modes.}
\vspace{5mm} \doublerulesep2pt
\begin{tabular}{ccccc}
\hline

\hline Modes&$\Gamma$( $10^{-16}$ GeV)&Br( $10^{-4}$)&Modes&Br( $\%$)
\\ \hline
$B^+\to
\bar{D}^0(2S)\ell^+{\nu_\ell}$&$2.05\pm0.61$&$5.1\pm1.5$&$B^+\to
\bar{D}^0(1S)\ell^+{\nu_\ell}$&$1.4\sim2.2$ \cite{B0D-}\\
$B^0\to D^-(2S)\ell^+{\nu_\ell}$&$1.95\pm0.54$&$4.5\pm1.3$&$B^0\to D^-(1S)\ell^+{\nu_\ell}$& $1.3\sim2.0$  \cite{B0D-}\\
$B_s^0\to D_s^-(2S)\ell^+{\nu_\ell}$&$4.3\pm1.2$&$9.9\pm2.7$&$B_s^0\to D_s^-(1S)\ell^+{\nu_\ell}$&$1.4\sim1.7$ \cite{BsDs}
\\
$B_c^+\to B^0(2S)\ell^+{\nu_\ell}$&$0.0171\pm0.0086$&$0.0120\pm0.0060$&$B_c^+\to B^0(1S)\ell^+{\nu_\ell}$&$0.090\sim0.11$\\
$B_c^+\to B_s^0(2S)\ell^+{\nu_\ell}$&$0.052\pm0.023$&$0.037\pm0.016$&$B_c^+\to B_s^0(1S)\ell^+{\nu_\ell}$&$1.2\sim1.6$\\
\hline

\hline
\end{tabular}
\end{table*}

In figure \ref{figform}, as an example, we plot the form factors
of the decay $B^0\to D^-(2S)\ell^+{\nu_\ell}$, as a comparison,
and we also show the form factors of $B^0\to
D^-(1S)\ell^+{\nu_\ell}$ in the same method, where
$t=(P-P_f)^2=M^2+M_f^2-2ME_f$ and $t_m$ is the maximum of $t$. One
can see that the values of form factors of $B^0\to D^-(2S)$ are
much smaller than that of $B^0\to D^-(1S)$, and they have
different shapes in figures. The reasons of these differences,
especially the different shape, are mainly caused by the different
wavefunctions of $1S$ and $2S$ states shown in figure
\ref{wavefunction}; the numerical values of the wavefunctions for
$D^-(1S)$ are positive and decrease along with the increased
momentum $|q|$, while there is a node structure in $D^-(2S)$
wavefunction, after the node, the wavefunctions are negative, and
whose negative values increase along with momentum $|q|$. This
negative part wavefunctions are responsible for the small decay
width and special shape of form factors; similar behaviors of the
form factors are also obtained in \cite{Ebert2,leibovich}.

In TABLE~\ref{widthbr}, we show the semi-leptonic decay widths and
branching ratios with final mesons being $2S$ states, also the
ones of corresponding $1S$ states in the same method for
comparison. With the same initial particle and same CKM matrix
element values, the branching ratios of the $2S$ channels are much
smaller than the ones of the $1S$ channels. This can be understood
by the differences of phase spaces and the node structure of
wavefunctions of 2$S$ state. The uncertainties of masses and decay
widths shown in TABLE~\ref{masses} and TABLE~\ref{widthbr} are
very large, some of them are almost 30\%.
 The large uncertainties not only come from the
uncertainties of phase spaces, but also from the variation of the
node of the $2S$ wave function, that means a little change of node
location will result in large uncertainties.

Although compared with ground-state cases, the production ratios
of $2S$ states are very small, the branching ratios of $B^0\to
D^-(2S)\ell^+{\nu_\ell}$ and $B^+\to
\bar{D}^0(2S)\ell^+{\nu_\ell}$ around $10^{-3}\sim10^{-4}$ are
very considerable. They are much larger than that of some rare
decay modes and are accessible in the current $B$ decay data. For
the channel of $B^0_s\to D_s^-(2S)\ell^+{\nu_\ell}$, the branching
ratio of order $10^{-3}$ can also be accessible in the near
future. But for the case of $B^+_c\to {B}^0(2S)\ell^+{\nu_\ell}$,
due to the small $V_{cd}$ and the phase space, we obtain narrow
decay width and small branching ratio. For the decay $B^+_c\to
{B}^0_s(2S)\ell^+{\nu_\ell}$, $V_{cs}$ is large, while the phase
space is very small, so the decay rate is small too. We also point
out that small phase space and the node structure result in almost
50\% uncertainties for the channels of $B^+_c\to
{B}^0(2S)\ell^+{\nu_\ell}$ and $B^+_c\to
{B}^0_s(2S)\ell^+{\nu_\ell}$.

\subsection{Strong decays}

These heavy-light $2S$ states can be detected experimentally
through their strong decays. In the previous letter, we have
calculated the strong decays of some $2S$ mesons \cite{zhangwang}.
But the predicted mass of $D^0(2S)$ is smaller than the
experimental data of $D(2550)^0$, which we have analyzed the
reasons in the introduction, so we got narrow widths of $B_q(2S)$
and $D_q(2S)$ in terms of their OZI-allowed strong decays. In this
paper, taking new parameters, we re-calculate the strong decays of
$B_q(2S)$ and $D_q(2S)$ with large masses. For example, we
calculate the strong decay $D^{*-}\to\bar D^0\pi^-$ by the
low-energy approximation, as well as the impulse approximation,
and obtain the strong coupling constant
$G_{D^*D\pi}$=19.8(18.3)\footnote[1]{The first value come from the
low-energy approximation, and the value in parentheses comes from
the impulse approximation.}, which is very closed to the
experimental value $G_{D^*D\pi}=17.9\pm0.3\pm1.9$ \cite{experi1}
and the result of quenched lattice QCD calculation
$G_{D^*D\pi}=18.8\pm2.3^{+1.1}_{-2.0}$ \cite{latticeQCD}. For
$B^{*0}\to B^+\pi^-$, which is phase space suppressed, but as one
more test of our method, similar to \cite{cdroberts1,bbpi,bbpi1},
we consider final $\pi$ as a soft-pion, and obtain the strong
coupling constant $G_{B^*B\pi}$=52.7(37.2)$^1$, the impulse
approximation result $37.2$ is close to the same method result
$G_{B^*B\pi}=30.0^{+3.2}_{-1.4}$ in \cite{cdroberts1}, but come
with a discrepancy may come from the further instantaneous
approach. Then, for the first radial excited states, we calculate
the transition matrix elements of the strong decay channels,
$0^-(2S)\to1^-0^-,0^+0^-$, and obtain the strong coupling
constants$^1$:
\begin{eqnarray}\label{strongconstant}
&&G_{\bar D^0(2S)\bar D^{*0}\pi^0}=3.71\pm0.10(3.96\pm0.34),\ \ G_{\bar D^0(2S)D^{*-}\pi^+}=5.24\pm0.25(5.62\pm0.45), \nonumber\\
&&G_{\bar D^0(2S)\bar D^{*0}_0\pi^0}=0.82\pm0.13(0.340\pm0.032)~{\rm GeV},\ \ G_{\bar D^0(2S)D^{*-}_0\pi^+}=1.10\pm0.33(0.58\pm0.15)~{\rm GeV}, \nonumber\\
&&G_{D^-(2S)D^{*-}\pi^0}=3.71\pm0.10(3.91\pm0.22),\ \ G_{D^-(2S)\bar D^{*0}\pi^-}=5.27\pm0.22(5.69\pm0.60), \nonumber\\
&&G_{D^-(2S)D^{*-}_0\pi^0}=0.70\pm0.12(0.40\pm0.10)~{\rm GeV},\ \ G_{D^-(2S)\bar D^{*0}_0\pi^-}=0.92\pm0.35(0.63\pm0.22)~{\rm GeV},\nonumber\\
&&G_{D_s^-(2S)\bar D^{*0}K^-}=7.76\pm0.65(6.76\pm0.50),\ \ G_{D_s^-(2S) D^{*-}\bar K^0}=7.77\pm0.60(6.68\pm0.47),\nonumber\\
&&G_{B^+(2S)B^{*+}\pi^0}=9.93\pm0.25(11.02\pm0.31),\ \ G_{B^+(2S)B^{*0}\pi^+}=14.04\pm0.33(15.51\pm0.62),\nonumber\\
&&G_{B^+(2S)B^{*+}_{0}\pi^0}=1.32\pm0.10(1.62\pm0.12)~{\rm GeV},\ \ G_{B^+(2S)B^{*0}_0\pi^+}=2.01\pm0.25(2.22\pm0.30)~{\rm GeV},\nonumber\\
&&G_{B^0(2S)B^{*0}\pi^0}=9.89\pm0.30(11.02\pm0.39)~,\ \ G_{B^0(2S)B^{*+}\pi^-}=14.07\pm0.34(15.72\pm0.56),\nonumber\\
&&G_{B^0(2S)B^{*0}_0\pi^0}=1.33\pm0.11(1.62\pm0.12)~{\rm GeV},\ \ G_{B^0(2S)B^{*+}_0\pi^-}=1.88\pm0.22(2.34\pm0.25)~{\rm GeV},\nonumber\\
&&G_{B_s(2S)B^{*+}K^-}=20.09\pm0.44(17.81\pm0.68),\ \
G_{B_s(2S)B^{*0} \bar K^0}=20.22\pm0.28(17.36\pm0.42).
\end{eqnarray}
Tables~\ref{strongww} and \ref{totalstrong} show the strong decay
widths. Comparing equation (\ref{strongconstant}) with table
\ref{strongww}, we find that the uncertainties of strong decay
widths are very large, even if there are small uncertainties of
strong coupling constants. This indicates that the predicted decay
widths are very sensitive to the mass (or kinematic range). The
two methods adopted in this paper, i.e. the low-energy
approximation and the the impulse approximation, obtained similar
strong decay widths, except some of the channels with $P$-wave
involved, which have large discrepancies. As one can see, these
$P$-wave-involved channels have much smaller phase spaces than
other channels, which show that these two methods give different
results in small phase space. And we pointed out one more time
that the large uncertainties in table~\ref{strongww} and
\ref{totalstrong} show that the decay widths are very sensitive to
the node structures of $2S$ states.

\begin{table*}\footnotesize
\centering  \caption{\label{strongww}\small The decay widths (in
unit of MeV) of strong decay modes. $\Gamma_1$ and $\Gamma_2$ come
from the impulse approximation and the low-energy approximation,
respectively.} \vspace{5mm} \doublerulesep2pt
\begin{tabular}{c|c|c|c|c|c}
\hline

\hline Modes&$\Gamma_1$&$\Gamma_2$&Modes&$\Gamma_1$&$\Gamma_2$
\\ \hline
$\bar D^0(2S)\to \bar D^{*0}\pi^0$&$15.2\pm6.9$&$14.1\pm7.5$&$\bar
D^0(2S)\to D^{*-}\pi^+$&$31\pm10$&$27\pm15$\\
$\bar D^0(2S)\to \bar
D^{*}_{0}(2400)^0\pi^0$&$0.182\pm0.058$&$0.83\pm0.30$&$\bar D^0(2S)\to
D^{*}_{0}(2400)^-\pi^+$&$0.35\pm0.20$&$1.39\pm0.57$
\\
$D^-(2S)\to D^{*-}\pi^0$&$16.4\pm6.6$&$14.5\pm7.5$&$D^-(2S)\to
\bar D^{*0}\pi^-$&$33\pm10$&$30\pm15$\\$D^-(2S)\to
D^{*}_{0}(2400)^-\pi^0$&$0.21\pm0.10$&$0.62\pm0.30$&$D^-(2S)\to
 \bar
D^{*}_{0}(2400)^0\pi^-$&$0.52\pm0.32$&$1.10\pm0.62$
\\
$D_s^-(2S)\to \bar
D^{*0}K^-$&$19\pm20$&$25\pm26$&$D_s^-(2S)\to D^{*-}\bar
K^0$&$17\pm18$&$24\pm25$
\\
$B^+(2S)\to B^{*+}\pi^0$&$30.1\pm4.0$&$24.1\pm6.0$&$B^+(2S)\to
B^{*0}\pi^+$&$59\pm13$&$48\pm12$\\
$B^+(2S)\to
B^{*+}_{0}\pi^0$&$0.85\pm0.25$&$0.45\pm0.15$&$B^+(2S)\to
B^{*0}_{{0}}\pi^+$&$1.61\pm0.41$&$1.04\pm0.31$
\\
$B^0(2S)\to B^{*0}\pi^0$&$30.1\pm4.8$&$24.0\pm4.4$&$B^0(2S)\to
B^{*+}\pi^-$&$60\pm13$&$48\pm12$\\
$B^0(2S)\to B^{*0}_{0}\pi^0$&$0.85\pm0.24$&$0.46\pm0.15$&
$B^0(2S)\to B^{*+}_{0}\pi^-$&$1.81\pm0.50$&$0.91\pm0.31$
\\
$B_s(2S)\to B^{*+}
K^-$&$42\pm18$&$55\pm19$&$B_s(2S)\to B^{*0} \bar K^0$&$41\pm18$&$54\pm19$
\\ \hline

\hline
\end{tabular}
\end{table*}

We find that there are two dominant OZI-allowed decay channels for
all the heavy-light $2S$ states in table~\ref{strongww}, both of
which are the case of $0^-(2S)\to 1^-0^-$. Other available
OZI-allowed strong decay channels ($0^-(2S)\to 0^+0^-,~2^+0^-$)
include heavy $P$-wave meson ($0^+$ or $2^+$) in the final states,
e.g., $\bar D^0(2S)\to \bar D^{*}_0(2400)^0\pi^0$, where
$D^{*}_0(2400)^0$ is a scalar ($0^+$) meson. Compared with the
decays whose final states are all $S$-wave meson ($1^-$ and
$0^-$), that includes $P$-wave meson in the final state and has a
small decay width. In table~\ref{strongww} we do not show the
channels that involve heavy tensor meson ($J^P=2^+$), which has an
ignorable decay width caused by the extremely narrow kinematic
range. There are not enough experimental data for the scalars
mesons ($J^P=0^+$). For example, $\bar D^{*}_0(2400)^0$ has been
confirmed, but $D^{*}_0(2400)^-$ is missing. So we choose the same
mass for $\bar D^{*0}_0$ and $D^{*-}_0$, i.e. $M_{\bar
D^{*}_0(2400)^0}=2.296\pm0.095$ GeV, and the same mass for
$B^{*0}_0$ and $B^{*+}_0$, i.e. $M_{B^{*0}_0}=5.660\pm0.266$ GeV.
For the decays of $D_s(2S)$ and $B_s(2S)$, the channels involving
$P$-wave state in the final states are kinematic forbidden.

Since we have obtained higher masses for heavy-light $2S$ states
than the ones of the previous letter \cite{zhangwang}, we obtain
much broader widths, but the estimated full width of $\bar
D^0(2S)$ is still narrower than the experimental data of
$D(2550)^0$. Our prediction is $\Gamma_{\bar
D^0(2S)}=47\pm17(43\pm23)$ MeV, while from data is $130\pm12\pm13$
MeV \cite{2550}. The new state $D(2550)^0$ should be the state
$D^0(2S)$ around the $2540$ MeV, since other states can be ruled
out by the masses or decay modes. For example, $P$-wave $D_{sJ}^*$
states whose masses are closed to $2540$ MeV have strong decay
productions including $K$ or $D_s$ but not $D^*\pi$. The $P$-wave
$D^*_J$ states, which have masses about 100 MeV lower than $2540$
MeV, have strong decays different from $0^-(2S)$. For $D$-mesons,
which are $1^-(2S)$, $1^-(2D)$, or the mixing of $1^-(2S)$ and
$1^-(2D)$, their masses should be higher than 2540 MeV and we
already have the candidate $D^*(2600)$ \cite{2550} that have more
strong decay channels: $0^-1^-$, $0^-0^-$, $1^-1^-$, $1^-0^+$.
Therefore, the new state $D(2550)^0$ is the state $D^0(2S)$. The
discrepancy between the theoretical and experimental result comes
from that we only considered the dominant OZI-allowed strong
decays. Finally, we should point out that the experimental data
may vary along with new more precise detections.

\begin{table}\footnotesize
\centering  \caption{\label{totalstrong}\small The estimated full widths (in unit of
MeV) of the first radial excited states.
The first and second value (in brackets) come from low-energy approximation and impulse approximation, respectively.} \vspace{5mm}
\doublerulesep2pt
\begin{tabular}{cccccc}
 \hline

\hline $\bar
D^0(2S)$&$D^-(2S)$&$D_s^-(2S)$&$B^0(2S)$&$B^+(2S)$&$B_s(2S)$
\\ \hline
$43\pm23(47\pm17)$&$46\pm23(50\pm17)$&$49\pm51(36\pm38)$&$74\pm17(91\pm18)$&$73\pm19(93\pm19)$&$109\pm38(83\pm36)$
\\  \hline

\hline
\end{tabular}
\end{table}

\subsection{Product of semi-leptonic decay ratio and cascaded strong decay ratio}

We have calculated $B_{q'}$ semi-leptonic decay to $B_q(2S)$ and
$D_q(2S)$, where all the states are on mass shells; we also
calculated the main strong decays of $B_q(2S)$ and $D_q(2S)$,
e.g., we obtain the branching ratio of the semi-leptonic decay
$Br(B^+\to\bar D^0(2S)\ell^+{\nu_\ell})=(5.1\pm1.5)\times10^{-4}$,
the strong decay branching ratio $Br(\bar D^0(2S)\to
D^{*-}\pi^+)=\frac{\Gamma_{\bar D^0(2S)\to
D^{*-}\pi^+}}{\Gamma_{\bar D^0(2S)}}=0.63\pm0.48(0.66\pm0.15)$ and
$Br(\bar D^0(2S)\to \bar D^{*0}\pi^0)=\frac{\Gamma_{\bar
D^0(2S)\to D^{*0}\pi^0}}{\Gamma_{\bar
D^0(2S)}}=0.33\pm0.24(0.32\pm0.11)$. To reduce the influence of
the discrepancies caused by the theoretical strong decay widths,
and the ground $1S$ states $D^{(*)}$, $B^{(*)}$, and their decays
that are well known in the experiment, we multiply the branching
ratio of the semi-leptonic decay with the strong decay branching
ratio, which show the ability of experiment to detect the missing
$2S$ states, but we ignore the reconstructed efficiencies of
events in experiment, and the products of ratios are
\begin{eqnarray}\label{br1}
&&Br(B^+\to\bar D^0(2S)\ell^+{\nu_\ell})\times Br(\bar D^0(2S)\to
D^{*-}\pi^+)\approx[3.2\pm2.6(3.4\pm1.9)]\times10^{-4},\nonumber\\
&&Br(B^+\to\bar D^0(2S)\ell^+{\nu_\ell})\times Br(\bar D^0(2S)\to \bar
D^{*0}\pi^0)\approx[1.7\pm1.4(1.6\pm1.0)]\times10^{-4},
\end{eqnarray}
\begin{eqnarray}\label{br2}
Br(B^0\to D^-(2S)\ell^+{\nu_\ell})\times Br(D^-(2S)\to
\bar
D^{*0}\pi^-)\approx[3.0\pm2.3(3.0\pm1.6)]\times10^{-4},\nonumber\\
Br(B^0\to D^-(2S)\ell^+{\nu_\ell})\times Br(D^-(2S)\to
 D^{*-}\pi^0)\approx[1.4\pm1.1(1.5\pm0.9)]\times10^{-4},
\end{eqnarray}
\begin{eqnarray}\label{br3}
Br(B_s^0\to D_s^-(2S)\ell^+{\nu_\ell})\times
Br(D_s^-(2S)\to \bar D^{*0}K^-)\approx
[ 5.1\pm7.6(5.2\pm7.9)]\times10^{-4},\nonumber\\
Br(B_s^0\to D_s^-(2S)\ell^+{\nu_\ell})\times
Br(D_s^-(2S)\to D^{*-}\bar K^0)\approx
[4.8\pm7.3(4.7\pm7.1)]\times10^{-4},
\end{eqnarray}
\begin{eqnarray}\label{br4}
Br(B_c^+\to B^0(2S)\ell^+{\nu_\ell})\times Br(B^0(2S)\to
B^{*+}\pi^-)
\approx[0.79\pm0.47(0.79\pm0.46)]\times10^{-6},\nonumber\\
Br(B_c^+\to B^0(2S)\ell^+{\nu_\ell})\times Br(B^0(2S)\to
B^{*0}\pi^0) \approx[0.39\pm0.23(0.40\pm0.22)]\times10^{-6},
\end{eqnarray}
\begin{eqnarray}\label{br5}
Br(B_c^+\to B_s^0(2S)\ell^+{\nu_\ell})\times
Br(B_s(2S)\to B^{*+}
K^-)\approx[1.9\pm1.5(1.9\pm1.4)]\times10^{-6},\nonumber\\
Br(B_c^+\to B_s^0(2S)\ell^+{\nu_\ell})\times
Br(B_s(2S)\to B^{*0} \bar
K^0)\approx[1.8\pm1.5(1.8\pm1.4)]\times10^{-6}.
\end{eqnarray}
The decays of $B^+$ and $B^0$ have ratios of order $10^{-4}$,
which can be analyzed with current data at B-factories. The decay
of $B_s$ that has ratios of order $10^{-4}$ may be observed in the
future, while the $B_c$ decay ratio of order $10^{-6}$ is hard to
reach experimentally.

In summary, we have studied the productions of $D_q(2S)$ and
$B_q(2S)$ in the exclusive semi-leptonic $B_{q'}$ decays and the
strong decays of $D_q(2S)$ and $B_q(2S)$. Some of these decays
have the branching ratios of order $10^{-4}$, which could be
measured currently in experiments. For examples, the ratios
$Br(B^+\to\bar D^0(2S)\ell^+{\nu_\ell})\times Br(\bar D^0(2S)\to
\bar D^{*}\pi)\approx[ 4.9\pm4.0(5.0\pm2.9)]\times10^{-4}$ and
$Br(B^0\to D^-(2S)\ell^+{\nu_\ell})\times Br(D^-(2S)\to \bar
D^{*}\pi)\approx[4.4\pm3.4(4.5\pm2.5)]\times10^{-4}$ are
relatively large in $B$ decays, which could be detected by the two
current $B$-factories. For $D_s(2S)$, the ratio $Br(B_s^0\to
D_s^-(2S)\ell^+{\nu_\ell})\times Br(D_s^-(2S)\to \bar D^{*}\bar
K)\approx [9.9\pm14.9(9.9\pm15.0)]\times10^{-4}$ is also not
small, which will be reached in the future. We have also given the
strong coupling constants of $D_q(2S)$ and $B_q(2S)$, which maybe
observed experimentally. Although similar to other models'
results, our calculation also gave a smaller full decay width than
the experimental data, but the recent detected $D(2550)^0$ is very
likely the ${D}^0(2S)$ state.

 \noindent
{\Large \bf Acknowledgements} This work was supported in part by
the National Natural Science Foundation of China (NSFC) under
Grant numbers~10875032 and 11175051.

\appendix{
\section{Instantaneous BS Equation}

In this section, we briefly review the BS equation and its
instantaneous one, i.e., the Salpeter equation.

The BS equation is read as \cite{BS}
\begin{equation}
(\not\!{p_{1}}-m_{1})\chi(q)(\not\!{p_{2}}+m_{2})=
i\int\frac{d^{4}k}{(2\pi)^{4}}V(P,k,q)\chi(k)\;, \label{eq1}
\end{equation}
where $\chi(q)$ is the BS wavefunction, $V(P,k,q)$ is the
interaction kernel between the quark and anti-quark, and $p_{1}
and p_{2}$ are the momenta of the quark 1 and anti-quark 2.

We divide the relative momentum $q$ into two parts, i.e.,
$q_{\parallel}$ and $q_{\perp}$:
$$q^{\mu}=q^{\mu}_{\parallel}+q^{\mu}_{\perp}\;,$$
$$q^{\mu}_{\parallel}\equiv (P\cdot q/M^{2})P^{\mu}\;,\;\;\;
q^{\mu}_{\perp}\equiv q^{\mu}-q^{\mu}_{\parallel}\;.$$

In the instantaneous approach, the kernel $V(P,k,q)$ takes the
simple form\cite{Salp}:
$$V(P,k,q) \Rightarrow V(|\vec k-\vec q|)\;.$$

Let us introduce the notations $\varphi_{p}(q^{\mu}_{\perp})$ and
$\eta(q^{\mu}_{\perp})$ for a three-dimensional wavefunction as
follows:
$$
\varphi_{p}(q^{\mu}_{\perp})\equiv i\int
\frac{dq_{p}}{2\pi}\chi(q^{\mu}_{\parallel},q^{\mu}_{\perp})\;,
$$
\begin{equation}
\eta(q^{\mu}_{\perp})\equiv\int\frac{dk_{\perp}}{(2\pi)^{3}}
V(k_{\perp},q_{\perp})\varphi_{p}(k^{\mu}_{\perp})\;. \label{eq5}
\end{equation}
Then, the BS equation can be rewritten as
\begin{equation}
\chi(q_{\parallel},q_{\perp})=S_{1}(p_{1})\eta(q_{\perp})S_{2}(p_{2})\;.
\label{eq6}
\end{equation}
The propagators of the two constituents can be decomposed as:
\begin{equation}
S_{i}(p_{i})=\frac{\Lambda^{+}_{ip}(q_{\perp})}{J(i)q_{p}
+\alpha_{i}M-\omega_{i}+i\epsilon}+
\frac{\Lambda^{-}_{ip}(q_{\perp})}{J(i)q_{p}+\alpha_{i}M+\omega_{i}-i\epsilon}\;,
\label{eq7}
\end{equation}
with
\begin{equation}
\omega_{i}=\sqrt{m_{i}^{2}+q^{2}_{_T}}\;,\;\;\;
\Lambda^{\pm}_{ip}(q_{\perp})= \frac{1}{2\omega_{ip}}\left[
\frac{\not\!{P}}{M}\omega_{i}\pm
J(i)(m_{i}+{\not\!q}_{\perp})\right]\;, \label{eq8}
\end{equation}
where $i=1, 2$ for quark and anti-quark, respectively,
 and
$J(i)=(-1)^{i+1}$.

We introduce the notations $\varphi^{\pm\pm}_{p}(q_{\perp})$ as
\begin{equation}
\varphi^{\pm\pm}_{p}(q_{\perp})\equiv
\Lambda^{\pm}_{1p}(q_{\perp})
\frac{\not\!{P}}{M}\varphi_{p}(q_{\perp}) \frac{\not\!{P}}{M}
\Lambda^{{\pm}}_{2p}(q_{\perp})\;. \label{eq10}
\end{equation}

With contour integration over $q_{p}$ on both sides of equation
(\ref{eq6}), we obtain
$$
\varphi_{p}(q_{\perp})=\frac{
\Lambda^{+}_{1p}(q_{\perp})\eta_{p}(q_{\perp})\Lambda^{+}_{2p}(q_{\perp})}
{(M-\omega_{1}-\omega_{2})}- \frac{
\Lambda^{-}_{1p}(q_{\perp})\eta_{p}(q_{\perp})\Lambda^{-}_{2p}(q_{\perp})}
{(M+\omega_{1}+\omega_{2})}\;,
$$
and the full Salpeter equation
$$
(M-\omega_{1}-\omega_{2})\varphi^{++}_{p}(q_{\perp})=
\Lambda^{+}_{1p}(q_{\perp})\eta_{p}(q_{\perp})\Lambda^{+}_{2p}(q_{\perp})\;,
$$
$$(M+\omega_{1}+\omega_{2})\varphi^{--}_{p}(q_{\perp})=-
\Lambda^{-}_{1p}(q_{\perp})\eta_{p}(q_{\perp})\Lambda^{-}_{2p}(q_{\perp})\;,$$
\begin{equation}
\varphi^{+-}_{p}(q_{\perp})=\varphi^{-+}_{p}(q_{\perp})=0\;.
\label{eq11}
\end{equation}

For the different $J^{PC}$ (or $J^{P}$) states, we give the
general form of wavefunctions (we will talk about them in appendix
B). Reduce the wavefunctions by the last equation of (\ref{eq11}),
and then solve the first and second equations of (\ref{eq11}) to
obtain the wavefunctions and mass spectrum. We have discussed the
solution of the Salpeter equation in detail in \cite{mass,w1}.

The normalization condition for the BS wavefunction is
\begin{equation}
\int\frac{q_{_T}^2dq_{_T}}{2{\pi}^2}Tr\left[\overline\varphi^{++}
\frac{{/}\!\!\!
{P}}{M}\varphi^{++}\frac{{/}\!\!\!{P}}{M}-\overline\varphi^{--}
\frac{{/}\!\!\! {P}}{M}\varphi^{--}\frac{{/}\!\!\!
{P}}{M}\right]=2P_{0}\;. \label{eq12}
\end{equation}

 In our model, the instantaneous interaction kernel $V$ is the Cornell
potential, which is the sum of a linear scalar interaction and a vector interaction:
\begin{equation}\label{vrww}
V(r)=V_s(r)+V_0+\gamma_{_0}\otimes\gamma^0 V_v(r)= \lambda
r+V_0-\gamma_{_0}\otimes\gamma^0\frac{4}{3}\frac{\alpha_s}{r}~,
\end{equation}
 where $\lambda$ is the string constant and $\alpha_s(\vec
q)$ is the running coupling constant. In order to fit the data of
heavy quarkonia, a constant $V_0$ is often added to the confine
potential. One can see that $V_v(r)$ diverges at $r=0$; we
introduce a factor $e^{-\alpha r}$ to avoid the divergence
\begin{equation}
V_s(r)=\frac{\lambda}{\alpha}(1-e^{-\alpha r})~,
~~V_v(r)=-\frac{4}{3}\frac{\alpha_s}{r}e^{-\alpha r}~.
\end{equation}\label{vsvv}
 It is easy to
know that when $\alpha r\ll1$, the potential becomes equation
(\ref{vrww}). In the momentum space and the center of mass system
of the bound state, the potential reads :
$$V(\vec q)=V_s(\vec q)
+\gamma_{_0}\otimes\gamma^0 V_v(\vec q)~,$$
\begin{equation}
V_s(\vec q)=-(\frac{\lambda}{\alpha}+V_0) \delta^3(\vec
q)+\frac{\lambda}{\pi^2} \frac{1}{{(\vec q}^2+{\alpha}^2)^2}~,
~~V_v(\vec q)=-\frac{2}{3{\pi}^2}\frac{\alpha_s( \vec q)}{{(\vec
q}^2+{\alpha}^2)}~,\label{eq16}
\end{equation}
where the running coupling constant $\alpha_s(\vec q)$ is
$$\alpha_s(\vec q)=\frac{12\pi}{33-2N_f}\frac{1}
{\log (a+\frac{{\vec q}^2}{\Lambda^{2}_{QCD}})}~.$$ We introduce a small
parameter $a$ to
avoid the divergence in the denominator. The constants $\lambda$, $\alpha$, $V_0$ and
$\Lambda_{QCD}$ are the parameters that characterize the potential. $N_f=3$ for $\bar bq$ (and $\bar cq$) system.

\section{Wavefunctions for different states}

We know that form factors of semi-leptonic decay and strong
coupling constants of strong decay are related to wavefunctions of
different states in section II and III. In this section, we give
the wavefunctions of the different states and
 obtain the form factors and strong coupling constants.

\noindent {\bf a). For $B_{q'}$ meson with quantum numbers
$J^{P}=0^{-}$}

The general form for the relativistic wavefunction of pseudoscalar
meson can be written as \cite{w1}
\begin{eqnarray}\label{aa01}
\varphi_{0^-}(\vec q)&=&\Big[f_1(\vec q){\not\!P}+f_2(\vec q)M+
f_3(\vec q)\not\!{q_\bot}+f_4(\vec q)\frac{{\not\!P}\not\!{q_\bot}}{M}\Big]\gamma_5,
\end{eqnarray}
where $M$ is the mass of the pseudoscalar meson, and $f_i(\vec q)$
are functions of $|\vec q|^2$. Due to the last two equations of
(\ref{eq11}): $\varphi_{0^-}^{+-}=\varphi_{0^-}^{-+}=0$, we have
\begin{eqnarray}\label{constrain}
f_3(\vec q)&=&\frac{f_2(\vec q)
M(-\omega_1+\omega_2)}{m_2\omega_1+m_1\omega_2},~~~
f_4(\vec q)=-\frac{f_1(\vec q)
M(\omega_1+\omega_2)}{m_2\omega_1+m_1\omega_2}.
\end{eqnarray}
where $m_1$ and $m_2$ and $\omega_1=\sqrt{m_1^{2}+\vec{q}^2}$,
$\omega_2=\sqrt{m_2^{2}+\vec{q}^2}$ are the masses and the
energies of
 quark and anti-quark in $B_{q'}$ mesons, $q_{_\bot}=q-(q\cdot P/M^2)P$, and $q_{\bot}^2=-|\vec q|^2$.
Then, there are only two independent unknown wavefunctions
$f_1(\vec q)$ and $f_2(\vec q)$ in equation (\ref{aa01}):
\begin{eqnarray}\label{aa012}
\varphi_{0^-}(\vec q)&=&\Big[f_1(\vec q){\not\!P}
+f_2(\vec q)M-f_2(\vec q)\not\!{q_\bot}
\frac{M(\omega_1-\omega_2)}{m_2\omega_1+m_1\omega_2}\nonumber\\
&&+f_1(\vec q){\not\!{q_{_{\bot}}}\not\!P}
\frac{\omega_1+\omega_2}{m_2\omega_1 +m_1\omega_2}\Big]\gamma_5.
\end{eqnarray}
The numerical values of radial wavefunctions $f_1$ and $f_2$ and
eigenvalue $M$ can be obtained by solving the first two Salpeter
equations in equation (\ref{eq11}).

 According to equation (\ref{eq10}), the relativistic positive wavefunction of
 pseudoscalar $0^-$ state in center of mass system can be written as \cite{w1}
\begin{eqnarray}\label{0-postive}
{\varphi}^{++}_{0^-}(\vec{q})=b_1
\left[b_2+\frac{\not\!{P}}{M}+b_3\not\!{q_{\bot}}
+b_4\frac{\not\!{q_{\bot}}\not\!{P}}{M}\right]{\gamma}_5,
\end{eqnarray}
where the $b_i$s ($i=1,~2,~3,~4$) are related to the original
radial wavefunction $f_i$, quark masses $m_i$, quark energy $w_i$
($i=1,~2$) and meson mass $M$:
$$b_1=\frac{M}{2}\left({f}_{1}(\vec{q})
+{f}_{2}(\vec{q})\frac{m_1+m_2}{\omega_1+\omega_2}\right),
b_2=\frac{\omega_1+\omega_2}{m_1+m_2}, b_3=-\frac{(m_1-m_2)}{m_1\omega_2+m_2\omega_1},
b_4=\frac{(\omega_1+\omega_2)}{(m_1\omega_2+m_2\omega_1)}.$$

\noindent {\bf b). For $B_{q}(2S)$ and $D_{q}(2S)$ mesons with quantum numbers
$J^{P}=0^{-}$}

Because the $2S$ state mesons have the same quantum numbers as
$B_{q'}$, the wavefunction of $2S$ state mesons are similar to
equation (\ref{0-postive}),
\begin{eqnarray}\label{final-2S}
{\varphi}^{++}_{_{P_f}}(\vec{q}_{f})=a_1
\left[a_2+\frac{\not\!{P_f}}{M_f}+a_3\not\!{q_{_{f\bot}}}
+a_4\frac{\not\!{q_{_{f\bot}}}\not\!{P_f}}{M_f}\right]{\gamma}_5,
\end{eqnarray}
$$a_1=\frac{M_f}{2}\left(f'_1(\vec{q}_f)
+f'_2(\vec{q}_f)\frac{m'_1+m'_2}{\omega'_1+\omega'_2}\right),
a_2=\frac{\omega'_1+\omega'_2}{m'_1+m'_2},
a_3=-\frac{(m'_1-m'_2)}{m'_1\omega'_2+m'_2\omega'_1},
a_4=\frac{(\omega'_1+\omega'_2)}{(m'_1\omega'_2+m'_2\omega'_1)}.$$
where $M_f$, $P_f$ and $f'_i(\vec q_f)$ are the mass, momentum and
the radial wavefunction of $2S$ state mesons, respectively. $m'_1,
m'_2$ and
$\omega'_1=\sqrt{m_1^{\prime2}+\vec{q}_f^2},\omega'_2=\sqrt{m_2^{\prime2}+\vec{q}_f^2}$
are the masses and the energies of
 quark and anti-quark in $2S$ state mesons, respectively.

According to the equations (\ref{matrix}), (\ref{0-postive}) and
(\ref{final-2S}), the form factors of $B_{q'}$ semi-leptonic
decays to $2S$ state mesons can be written as
 \begin{eqnarray}\label{f+}
&&f_+=\frac{1}{2}\int\frac{d^3q}{(2\pi)^3}\frac{4a_1b_1}{MM_f}
\left[(a_2b_2M_f+b_3q\cdot
r\cos\theta-a_4b_2E_f^2{\alpha}'_1-a_4b_4E_fq\cdot
r\cos\theta){\alpha}'_1\right.\nonumber\\
&&+a_3M_f(b_3q^2+E_f{\alpha}'_1-b_3q\cdot
r\cos\theta{\alpha}'_1)+M(1+a_4b_4q^2+a_4b_2E_f{\alpha}'_1-a_3M_f{\alpha}'_1)\nonumber\\
&&\left.-M(1-\frac{E_f}{M})(a_4b_2E_f-b_3E_f-a_3M_f+a_2b_4M_f+a_4b_4q\cdot
r\cos\theta)\right],
\end{eqnarray}
\begin{eqnarray}\label{f-}
&&f_-=\frac{1}{2}\int\frac{d^3q}{(2\pi)^3}\frac{4a_1b_1}{MM_f}
\left[(a_2b_2M_f+b_3q\cdot
r\cos\theta-a_4b_2E_f^2{\alpha}'_1-a_4b_4E_fq\cdot
r\cos\theta){\alpha}'_1\right.\nonumber\\
&&+a_3M_f(b_3q^2+E_f{\alpha}'_1-b_3q\cdot
r\cos\theta{\alpha}'_1)-M(1+a_4b_4q^2+a_4b_2E_f{\alpha}'_1-a_3M_f{\alpha}'_1)\nonumber\\
&&\left.+M(1+\frac{E_f}{M})(a_4b_2E_f-b_3E_f-a_3M_f+a_2b_4M_f+a_4b_4q\cdot
r\cos\theta)\right],
\end{eqnarray}
where $E_f=\sqrt{M_f^2+\vec r^2}$, $q\cdot r \equiv |\vec q \cdot
\vec r|$, $\theta$ is the angle between $\vec q$ and $\vec r$.

\noindent {\bf c). For $B^*_{q}$ and $D^*_{q}$ mesons with quantum numbers
$J^{P}=1^{-}$}

The relativistic positive wavefunction of $1^-$ state can be
written as
\begin{eqnarray}\label{postive-1}
{\varphi}_{1^-}^{++}(\vec{q}_{f1})&=&c_1\not\!{\varepsilon}^{(\lambda)}+c_2\not\!{\varepsilon}^{(\lambda)}\not\!{P_{f1}}
+c_3(\not\!{q_{f1\bot}}\not\!{\varepsilon}^{(\lambda)}-q_{f1\bot}\cdot{\varepsilon}^{(\lambda)})
+c_4(\not\!{P}_{f1}\not\!{\varepsilon}^{(\lambda)}\not\!{q_{f1\bot}}-\not\!{P}_{f1}q_{f1\bot}\cdot{\varepsilon}^{(\lambda)})
\nonumber\\
&&+q_{f1\bot}\cdot{\varepsilon}^{(\lambda)}(c_5+c_6\not\!{P}_{f1}+c_7\not\!{q_{f1\bot}}+c_8\not\!{q_{f1\bot}}\not\!{P}_{f1}),
\end{eqnarray}
where we first defined the parameter $n_i$ that is functions of
$f''_i$ ($1^-$ wave functions):
$$n_1=f''_{5}(\vec{q}_{f1})
-f''_{6}(\vec{q}_{f1})\frac{(\omega''_1+\omega''_2)}{(m''_1+m''_2)}, n_2=f''_{5}(\vec{q}_{f1})
-f''_{6}(\vec{q}_{f1})\frac{(m''_1+m''_2)}{(\omega''_1+\omega''_2)},$$
$$ n_3=f''_{3}(\vec{q}_{f1})
+f''_{4}(\vec{q}_{f1})\frac{(m''_1+m''_2)}{(\omega''_1+\omega''_2)}.$$
Then, we defined the parameters $c_i$ that are functions of
$f''_i$ and $n_i$:
$$c_1=\frac{M_{f1}}{2}n_1, c_2=-\frac{1}{2}\frac{(m''_1+m''_2)}{(\omega''_1+\omega''_2)}n_1,
c_3=\frac{M_{f1}}{2}\frac{(\omega''_2-\omega''_1)}{(m''_1\omega''_2+m''_2\omega''_1)}n_1,
c_4=\frac{1}{2}\frac{(\omega''_1+\omega''_2)}{(\omega''_1\omega''_2+m''_1m''_2-{q_{f1\bot}^{2}})}n_1,$$
 $$c_5=\frac{1}{2M_{f1}}\frac{m''_1+m''_2}{(\omega''_1\omega''_2+m''_1m''_2+{q_{f1\bot}^{2}})}(M_{f1}^2n_2+{q_{f1\bot}^{2}}n_3),
  c_6=\frac{1}{2M_{f1}^2}\frac{\omega''_1-\omega''_2}{(\omega''_1\omega''_2+m''_1m''_2+{q_{f1\bot}^{2}})}(M_{f1}^2n_2+{q_{f1\bot}^{2}}n_3),$$
$$c_7=\frac{n_3}{2M_{f1}}-\frac{f''_6(\vec{q}_{f1})M_{f1}}{(m''_1\omega''_2+m''_2\omega''_1)},
 c_8=\frac{1}{2M_{f1}^2}\frac{\omega''_1+\omega''_2}{m''_1+m''_2}n_3
 -f''_5(\vec{q}_{f1})\frac{\omega''_1+\omega''_2}{(m''_1+m''_2)(\omega''_1\omega''_2+m''_1m''_2-{q_{f1\bot}^{2}})}.$$

 According to equations (\ref{a08}), (\ref{final-2S}) and (\ref{postive-1}),
 the strong coupling constant for $0^-(2S)\to 1^-0^-$ can be written as
 \begin{eqnarray}\label{G1-0-}
&&G_{0^-(2S)\to 1^-0^-}=\frac{1}{f_X}\int\frac{d^3q_f}{(2\pi)^3}4a_1\left\{\frac{1}{M_f}\left[(a_4c_3M_f+a_3c_4M_{f1}^2)|\vec q_{f}|^2
-a_4c_2M_f\vec q_f\cdot \vec P_{f1}\right.\right.\nonumber\\
&&+a_3c_4(\vec q_f\cdot \vec P_{f1})^2-c_1(M_f+a_3\vec q_f\cdot \vec P_{f1})+c_6E_{f1}\alpha''_2(E_{f1}M_f-M_{f1}^2)
+a_3c_7|\vec q_{f}|^2E_{f1}\alpha''_2(M_f-E_{f1})\nonumber\\
&&a_4c_8|\vec q_{f}|^2E_{f1}\alpha''_2(E_{f1}M_f-M_{f1}^2)+(c_7 E_{f1}-a_4c_5E_{f1}+a_3c_4E_{f1}^2+a_4c_3M_f+a_3c_6E_{f1}M_f)\vec q_f\cdot \vec P_{f1}\alpha''_2 \nonumber\\
&&
-a_4c_8E_{f1}\alpha''_2(\vec q_f\cdot \vec P_{f1})^2+c_7E_{f1}\alpha_2^{\prime\prime2}|\vec P_{f1}|^2+a_3c_7\vec q_f\cdot \vec P_{f1}\alpha_2^{\prime\prime2}E_{f1}(M_f-E_{f1})\nonumber\\
&&+a_4c_8E_{f1}^2\alpha_2^{\prime\prime2}(M_f-E_{f1})+a_2(c_2M_{f1}^2+c_3(\vec q_f\cdot \vec P_{f1}+|\vec P_{f1}|^2\alpha''_2)\nonumber\\
&&+E_{f1}\alpha''_2(c_5(E_{f1}-M_f)+c_8E_{f1}(\vec q_f\cdot \vec P_{f1}+E_{f1}^2\alpha''_2-M_{f1}^2\alpha''_2)))
\left.\right]-\frac{E_{f1}|\vec q_f|}{M_f|\vec P_{f1}|}\left[a_2c_5(M_f-E_{f1})\right.\nonumber\\
&&+(a_4c_2+c_6)(M_{f1}^2-E_{f1}M_f)+a_4c_8|\vec q_f|^2(M_{f1}^2-E_{f1}M_f)+(a_4c_3+a_4c_5-c_7-a_2c_8E_{f1})\vec q_f\cdot \vec P_{f1}\nonumber\\
&&+a_4c_8|\vec q_f\cdot \vec P_{f1}|^2+(a_4c_3-c_7)E_{f1}^2\alpha''_2-a_2c_8E^3_{f1}\alpha''_2+(c_7-a_4c_3)M_{f1}^2\alpha''_2+a_2c_8E_{f1}M_{f1}^2\alpha''_2\nonumber\\
&&+a_4c_8\vec q_f\cdot \vec P_{f1}\alpha''_2E_{f1}(E_{f1}-M_{f1})+a_3(c_1(M_f-E_{f1})+(c_4E_{f1}-c_6M_f)\vec q_f\cdot \vec P_{f1}\nonumber\\
&&\left.\left.+c_4E_{f1}\alpha''_2|\vec P_{f1}|^2+c_7(E_{f1}-M_f)(|\vec q_f|^2+\vec q_f\cdot \vec P_{f1}\alpha''_2))
\right]\right\}
\end{eqnarray}
 where $f_{X}$ ($X$ is $\pi$ or $K$) is the decay constant for the light meson, $\alpha''_2=\frac{m''_2}{m''_1+m''_2}$.

 \noindent {\bf d). For $B^*_{q}$ and $D^*_{q}$ mesons with quantum numbers
$J^{P}=0^{+}$}

The relativistic positive energy wavefunction of $0^+$ can be
written as
\begin{equation}\label{postive-0+}
{\varphi}_{0^+}^{++}(\vec{q}_{f1})=c_1(\not\!{q_{f1\bot}}+c_2\frac{\not\!{P}_{f1}\not\!{q_{f1\bot}}}{M_{f1}}
+c_3+c_4\frac{\not\!{P}_{f1}}{M_{f1}}),
\end{equation}
where the parameters $c_i$ are the functions of $f''_1$ and
$f''_2$ ($0^+$ wavefunction), which are defined as
$$c_1=\frac{1}{2}\left(f''_{1}(\vec{q}_{f1})
+f''_{2}(\vec{q}_{f1})\frac{m''_1+m''_2}{\omega''_1+\omega''_2}\right),
c_2=\frac{\omega''_1+\omega''_2}{m''_1+m''_2},
c_3=q_{f1\bot}^2\frac{(\omega''_1+\omega''_2)}{m''_1\omega''_2+m''_2\omega''_1},
c_4=\frac{(m''_2\omega''_1-m''_1\omega''_2)}{(m''_1+m''_2)}.$$

 Using equations (\ref{a08}), (\ref{final-2S}) and (\ref{postive-0+}),
 the strong coupling constant for $0^-(2S)\to 0^+0^-$ can be written as
  \begin{eqnarray}\label{G0+0-}
  &&G_{0^-(2S)\to 0^+0^-}=\frac{1}{f_X}\int\frac{d^3q_f}{(2\pi)^3}\frac{-4a_1c_1}{M_{f1}}\left\{(a_4c_2E_{f1}M_f-a_3E_{f1}M_{f1}+a_3M_f M_{f1}-a_4c_2M^2_{f1})|\vec q_f|^2\right.\nonumber\\
  &&+(M_{f1}-a_4c_3M_{f1})\vec q_f\cdot \vec P_{f1}-a_4c_2|\vec q_f\cdot \vec P_{f1}|^2+c_4(E_{f1}M_f-M^2_{f1}+a_3M_f\vec q_f\cdot \vec P_{f1})\nonumber\\
  &&+|\vec P_{f1}|^2M_{f1}\alpha''_2+(a_4c_2E_{f1}+M_{f1}a_3)(M_f-E_{f1})\vec q_f\cdot \vec P_{f1}\alpha''_2
  +a_2(c_3M_{f1}(E_{f1}-M_f)\nonumber\\
  &&+c_2E_{f1}(\vec q_f\cdot \vec P_{f1}+|\vec P_{f1}|^2)\alpha''_2)\left.\right\}
  \end{eqnarray}
}

\end{document}